\documentclass[reprint,amsmath,amssymb, aps]{revtex4-1}

\usepackage{graphicx}% Include figure files
\usepackage{dcolumn}% Align table columns on decimal point
\usepackage{bm}% bold math

\usepackage{hyperref}                               % for links
\usepackage{textcomp}                               % numero sign
\usepackage{graphicx}
\graphicspath{{figures/}}

\usepackage{amsmath}
\usepackage{mathpazo}
\usepackage[makeroom]{cancel}
\usepackage{empheq}
\usepackage[most]{tcolorbox}

\newcommand{\subtxt}[1]{_{\textrm{#1}}}

\newcommand{\R}{\mathcal{R}}

\newcommand{\DC}{\subtxt{DC}}
\newcommand{\AC}{\subtxt{AC}}
\newcommand{\del}[1]{\Delta #1}

\newcommand{\deldc}[1]{\del{#1}\DC}

\newcommand{\delac}[1]{\del{#1}\AC}
\newcommand{\dT}{\del{T}}

\begin{document}

\title{Imaging the Thermalization of  Hot Carriers After Thermionic Emission Over a Polytype Barrier}

\author{Fabian K\"onemann}
\affiliation{IBM Research - Zurich, S\"aumerstrasse 4, 8803 R\"uschlikon, Switzerland}

\author{I-Ju Chen}
\affiliation{Solid State Physics and NanoLund, Lund University, S-221 00 Lund, Sweden}

\author{Sebastian Lehmann}
\affiliation{Solid State Physics and NanoLund, Lund University, S-221 00 Lund, Sweden}

\author{Claes Thelander}
\affiliation{Solid State Physics and NanoLund, Lund University, S-221 00 Lund, Sweden}

\author{Bernd Gotsmann}
\email{bgo@zurich.ibm.com}
\affiliation{IBM Research - Zurich, S\"aumerstrasse 4, 8803 R\"uschlikon, Switzerland  }

\date{\today}

\begin{abstract}
The thermalization of non-equilibrium charge carriers is at the heart of thermoelectric energy conversion. In nanoscale systems, the equilibration length can be on the order of the system size, leading to a situation where thermoelectric effects need to be considered as spatially distributed, rather than localized at junctions. The energy exchange between charge carriers and phonons is of fundamental scientific and technological interest, but their assessment poses significant experimental challenges.
We addressed these challenges by imaging the temperature change induced by Peltier effects in crystal phase engineered InAs nanowire (NW) devices. Using high-resolution scanning thermal microscopy (SThM), we have studied current-carrying InAs NWs, which feature a barrier segment of wurtzite (WZ) of varying length in a NW of otherwise zincblende (ZB) crystal phase. The energy barrier acts as a filter for electron transport around the Fermi energy, giving rise to a thermoelectric effect. We find that thermalization through electron-phonon heat exchange extends over the entire device. We analyze the temperature profile along a nanowire by comparing it to spatially dependent heat diffusion and electron thermalization models. We are able to extract the governing properties of the system, including the electron thermalization length of $223 \pm 9$\,nm, Peltier coefficient and Seebeck coefficient introduced by the barrier of $39 \pm 7$\,mV and $89 \pm 21$\,$\mu$V/K, respectively, and a thermal conductivity along the wire axis of $8.9 \pm 0.5$\,W/m/K. Finally, we compare two ways to extract the elusive thermal boundary conductance between NW and underlying substrate.
\end{abstract}

\maketitle

\begin{figure*}
    \centering
    \includegraphics[width=\textwidth]{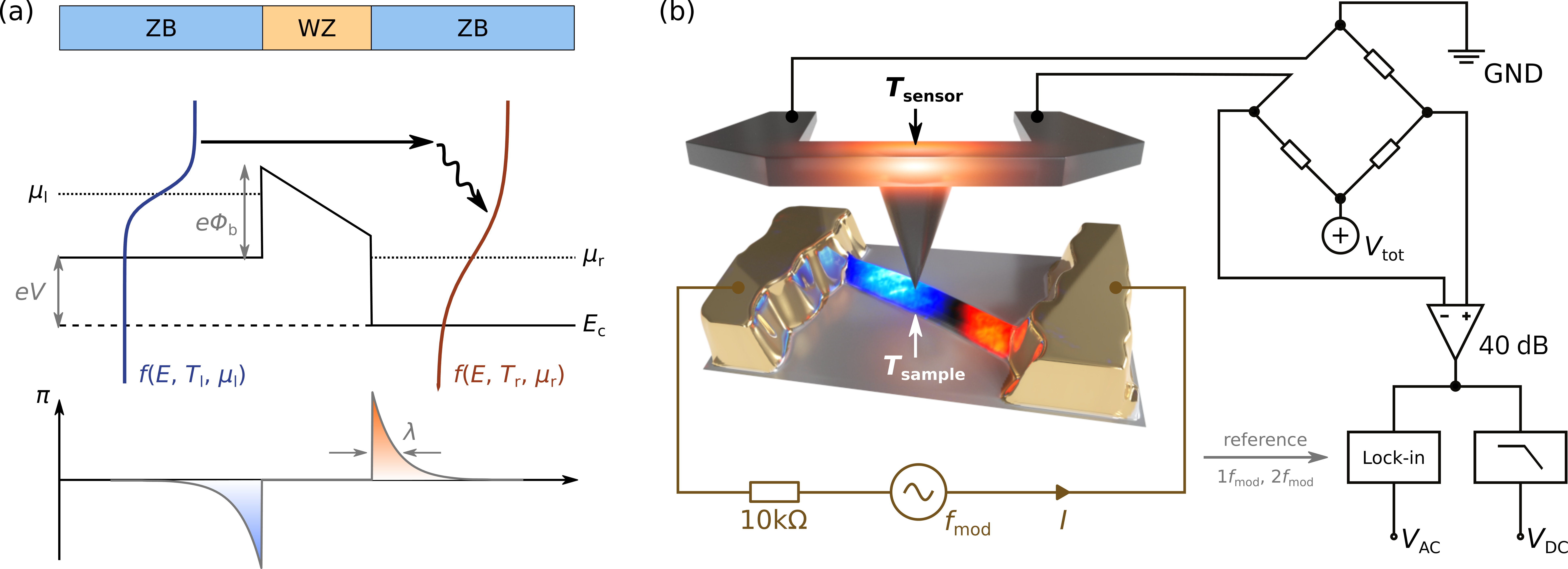}
    \caption{(a) Schematic representation of the thermionic emission effect that leads to Peltier heating / cooling. The conduction band energy $E\subtxt{c}$ has a positive offset $e\Phi\subtxt{b}$ in the WZ segment with respect to the surrounding ZB material. This gives rise to an energy barrier that acts as a filter for conduction electrons around the Fermi energy $\mu$. When a bias $V$ is applied across the WZ segment, such that the Fermi energy $\mu\subtxt{l}$ on the left is higher with respect to the Fermi energy $\mu\subtxt{r}$ on the right, only ``hot'' electrons from the high-energy tail of the left-side Fermi distribution can be emitted over the barrier and contribute to electrical conduction. They thermalize on the right side by scattering with a characteristic thermalization length $\Lambda$, widening up the electron and phonon distributions and thereby increasing the temperature on the right side. At the same time, the distributions on the left side are narrowed, meaning that the temperature decreases there. An approximately exponentially distributed heating/cooling power arises at the junctions. The electronic current density therefore does not only transport charge, but also heat across the barrier. (b) Schematic representation of the measurement technique. See introduction and experimental section for detailed explanations.} 
    \label{fig:figure_1}
\end{figure*}

\section{Introduction}

    Thermoelectric effects, such as Peltier and Seebeck effects, have been explored for solid state cooling and energy harvesting devices, respectively. However, the low intrinsic thermoelectric efficiency has been hampering a wide use of these effects in technology \cite{vining_inconvenient_2009}. One of the popular approaches to increase efficiency is to exploit energy filtering of charge carriers passing an energy barrier, an effect often referred to as thermionic emission~\cite{shakouri_heterostructure_1997, benenti_fundamental_2017, horio_numerical_1990, yang_numerical_1993, shakouri_a._thermoelectric_1998}. It is not straightforward to exploit this effect due to difficulties in creating high-quality (ideally atomically defined) interfaces between a reservoir material (semiconductor or metal) and a barrier region, and to control the chemical potential across such heterostructures. 
    The heterostructure between WZ and ZB crystal phase in InAs NWs are an interesting candidate for thermionic emission structures. In InAs and most other III-V NWs the crystal phase can be controlled during growth~\cite{caroff_controlled_2009, lehmann_general_2013, dick_control_2010, pan_controlled_2014}, and interfaces between the two phases are inherently atomically sharp. Such InAs NW crystal phase heterostructures have recently been studied in their transport properties, and it was found that the WZ phase has a positive conduction band offset of about 135\,meV compared to the ZB phase~\cite{chen_conduction_2017, belabbes_electronic_2012}. This property has successfully been exploited for creating quantum dot devices~\cite{nilsson_single-electron_2016, nilsson_parallel-coupled_2017, akopian_crystal_2010}.
    
    The study of thermoelectric effects in nanostructures is of relevance beyond energy conversion applications. With increasing miniaturization of electronic devices, thermoelectric effects have become important for operation for example in resistive\cite{wang_thermoelectric_2014, yalon_spatially_2017} and phase change\cite{suh_thermoelectric_2010, lee_impact_2012} memory devices or transistors~\cite{pop_energy_2010}, influencing power consumption and device failure. One of the scientific questions around such applications is related to the length scale over which thermalization occurs. In thermoelectric structures and devices on the macroscale, the generation of Peltier heat can safely be assumed to occur sharply at the interface between two materials. The equilibration of charge carriers passing a barrier, however, can be on the order of hundred nanometers and is therefore comparable or even larger than device dimensions. Therefore, on the nanoscale, the simplified picture of thermoelectricity must be refined. However, measurement techniques to probe thermalization have not been available. Thermal and thermoelectric effects in NWs and other nanostructures are generally difficult to quantify.~\cite{cahill_nanoscale_2002} 

    Here we apply an imaging technique to measure the temperature distribution of a thermoelectric nanowire device in operando. Using scanning thermal microscopy, we obtain such images with a spatial resolution of at least 10\,nm~\cite{menges_temperature_2016, koonemann_nanoscale_2018, konemann_thermal_2019}. The method has recently been used to study geometrically enhanced Peltier effects at a graphene constriction~\cite{harzheim_geometrically_2018}. 
    From the resulting temperature maps, we can extract thermal properties such as the thermal conductivity and the Peltier coefficient, and, importantly, the equilibration length of the charge carriers and a related density of Peltier heat production~\cite{lake_energy_1992, shakouri_a._thermoelectric_1998, chen_thermoelectric_2018}. Note that in this article, we use the term Peltier effect to refer both to heat pumping in response to an electrical current by thermionic emission over the barrier, and to heat pumping in response to an electrical current at the semiconductor/metal interfaces at the contacts.

    The sample system chosen here are InAs NWs of ZB phase with a segment of 40 to 160\,nm length of WZ, see \ref{fig:figure_1}a. The nanowires are on a silicon oxide on silicon substrate and contacted using nickel-gold metal contacts (\ref{fig:figure_1}b). Further details on device fabrication can be found in the appendix and in Ref. \cite{chen_conduction_2017}. The dominant carriers are n-type and the current-voltage relationship is approximately linear at ambient temperatures.

    We apply a scanning probe thermometry technique based on SThM. As shown in \ref{fig:figure_1}b, a sinusoidal alternating current with amplitude $I$ and frequency $f\subtxt{mod}$ is passed through the device. The device responds thermally with Peltier heating/cooling at $1f\subtxt{mod}$ and Joule heating at $2f\subtxt{mod}$. We denote these temperature responses with $\dT\subtxt{Peltier}$ and $\dT\subtxt{Joule}$, respectively. A silicon SThM probe with integrated resistive sensor is then raster-scanned across the sample. The sensor is thermally coupled to a spot on the sample surface through a sharp tip. The tip radius can be as small as 3\,nm~\cite{ryu_cho_sub-10_2017}. Changes in the sample surface temperature $T\subtxt{sample}$ influence the equilibrium temperature of the sensor $T\subtxt{sensor}$, which in turn changes the sensor element's electrical resistance. The sample topography is measured simultaneously as in standard contact mode atomic force microscopy (AFM).

\section{Results and Discussion}
    \begin{figure*}
        \centering
        \includegraphics[width=\textwidth]{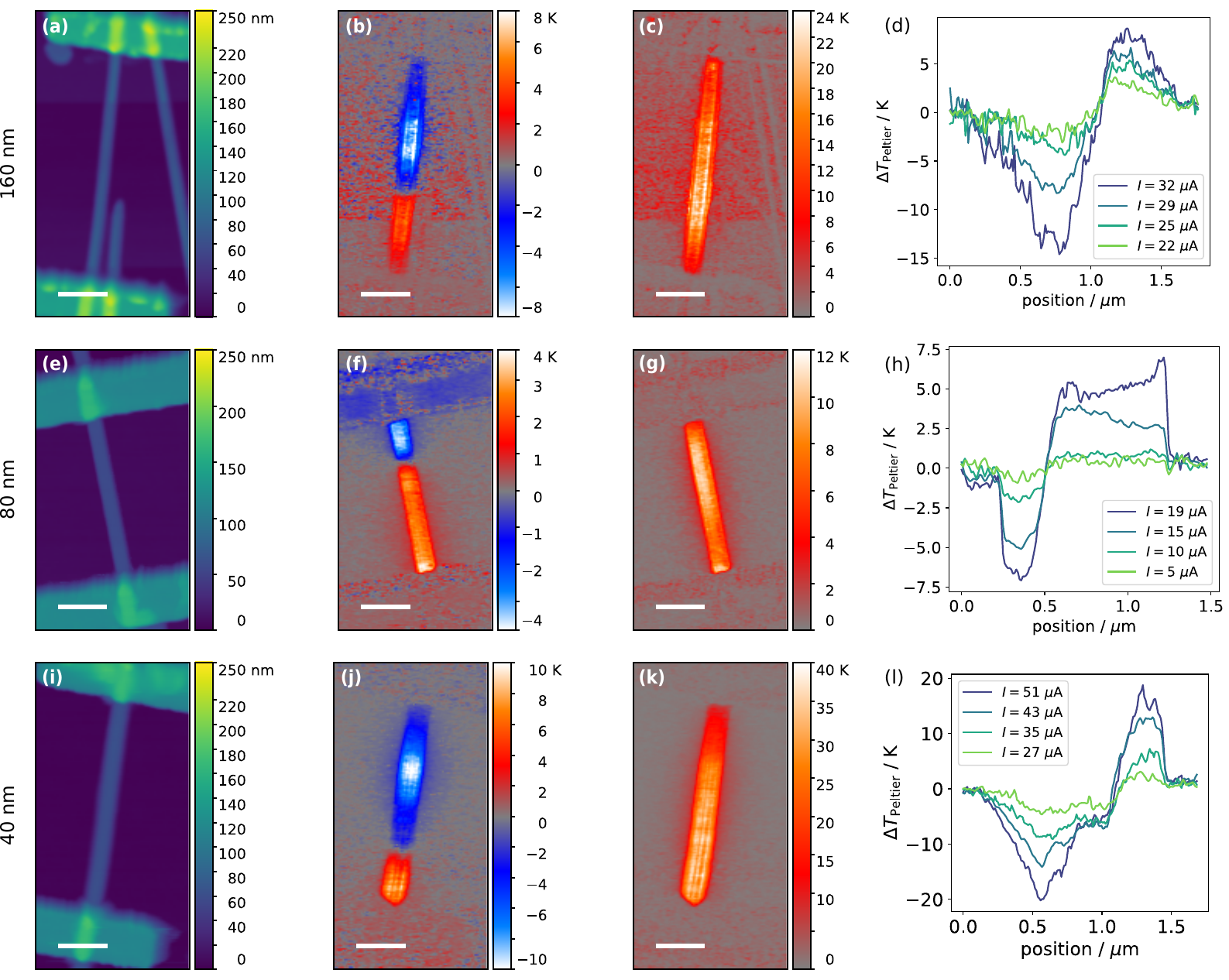}
        \caption{SThM results for three devices with different targeted WZ segment lengths: (a) AFM topography map for 160\,nm, (b) $\dT\subtxt{Peltier}$ map for 160\,nm, (c) $\dT\subtxt{Joule}$ map for 160\,nm, and (d) line profiles of $\dT\subtxt{Peltier}$ along the 160\,nm nanowire axis for different values of $I$. (e-h) The same depictions for 80\,nm. (i-l) For 40\,nm. Shown temperature 
        change maps always correspond to the highest value of $I$ stated in the legend of the associated profile plot. The locations of the barriers are assumed to roughly coincide with the transition between heating and cooling in the $\dT\subtxt{Peltier}$ maps. The profiles 
        have an averaging width of five pixels at a square pixel size of 10\,nm. For all scans, the horizontal direction as depicted is the fast scanning direction. The display ranges in the temperature maps has been limited to increase the contrast. All scale bars 300\,nm.}
        \label{fig:different_old_wires_compilation}
    \end{figure*}
    
    We recorded Joule and Peltier SThM signals of self-heated NW samples with target WZ segment lengths of 40\,nm, 80\,nm, and 160\,nm. Temperature maps were recorded for different modulation current amplitudes $I$. Exemplary plots of AFM topography, $\dT\subtxt{Peltier}$ and $\dT\subtxt{Joule}$ for each device are shown in \ref{fig:different_old_wires_compilation}. From these, line profiles of $\dT\subtxt{Peltier}$ along the nanowires were extracted, which are also shown in \ref{fig:different_old_wires_compilation}. Profiles for $\dT\subtxt{Joule}$ and data for additional values of $I$ are shown in the supporting information. All three devices show a clear $\dT\subtxt{Peltier}$ signal. The position of the WZ segment along the NWs between the electrodes can be identified to be in the region of sign-change of the Peltier signal. Let us first discuss the markedly different qualitative behavior of the three devices shown.
    
    For the 160\,nm device, the peaks of $\dT\subtxt{Peltier}$ are close to the transition between heating and cooling, which is steep with a linear gradient. This is the expected behaviour for heating and cooling power that is concentrated close to the crystal phase junctions. This suggests that $\dT\subtxt{Peltier}$ is dominated by the barrier Peltier effect. However, the positions of maximum and minimum temperature are spaced further apart than the length of the WZ segment. This is an indication that the release of Peltier heat does not occur sharply at the crystal phase junctions, but is spatially distributed away from the junctions into the ZB sections towards the contacts. We will analyze this in more detail below. We note that the lateral resolution of SThM is required to reveal these observations.
    %The temperature between these hot and cold spots and the thermalized metal electrodes decays as expected for thermal conduction along a one-dimensional wire on a substrate.
    
    For the 80\,nm device, the profiles look significantly different. Here, $\dT\subtxt{Peltier}$ stays more uniform along the ZB sections. For the highest $I$ of 19\,$\mu$A, we see a peak of $\dT\subtxt{Peltier}$ at the bottom contact rather than at the transition between heating and cooling.  
    Unlike the previous example, Peltier heating/cooling is not concentrated around the WZ segment region. In addition, a Peltier effect due to the metal-NW interface appears to come into play. The sign of the associated Peltier coefficients (or Seebeck coefficients) are opposite for the two effects, such that they add up constructively on either side of the device. We expect that the Seebeck coefficient of an energy barrier is always positive, while the Seebeck coefficient of an InAs NW is negative~\cite{karg_full_2014} as observed also by SThM~\cite{menges_temperature_2016, menges_local_2016}. The Seebeck coefficient of the metal electrodes is two orders of magnitude lower in magnitude and can be neglected~\cite{rosa_fundamentals_2012}. It appears like the barrier effect rolls off for high currents, while the contact effect becomes more dominant. This is manifested in the slope changing from decreasing towards the bottom contact to increasing towards the bottom contact when going from $I\subtxt{mod} = 15\,\mu$A to 18\,$\mu$A in \ref{fig:different_old_wires_compilation}h. We acknowledge at this point that the barrier emission effect is inherently non-linear in nature. Assuming a constant effective Peltier coefficient corresponds to averaging by linearization. We do not observe any signal at odd-order higher harmonics, indicating the absence of significant distortions. We therefore believe that the consequences of this simplification are marginal.
    The very flat temperature rise observed is remarkable, as it  indicates that the heating/cooling power is distributed on a length scale comparable to the system size, which we interpret as a sign of an electron thermalization length on the order of 100 nanometers. The sharp peak observed at the bottom contact for $I = 19\mu$A might be an artifact of the tip scratching the contact, or it might indicate the existence of a Shottky barrier at the contact.
    
    The data for the 40\,nm device shows a continuation of the trend. Here, the transition between heating and cooling is not the point of steepest gradient. The temperature profile is almost flat at the transition in the region of the WZ segment. The peaks of $\dT\subtxt{Peltier}$ appear to be separated from the barrier region. This indicates that the contact Peltier effect is now dominant compared to the thermionic emission effect. The peaks are however not directly at the contacts and we observe an asymmetric profile, which confirms that the barrier also has an influence in the 40\,nm WZ system.

    In summary, the first comparison of the NWs with different WZ segment lengths suggests the presence of two Peltier effects, one at the energy barrier and one at the NW-metal contacts. The relative weight of the resulting temperature change in the according device areas depends on the segment length and potentially also on the applied bias. The first and obvious explanation for this trend is that for short barrier lengths, the heating and cooling poles are closer to each other. Since the material separating the two poles is in all cases WZ InAs, shorter segments lead to smaller thermal resistances between the two, meaning that the heating and cooling sides can exchange heat more efficiently. Therefore the barrier Peltier effect has to be less dominant for shorter WZ segments if we assume that the Peltier coefficient is identical for all barrier lengths. However, the shape of the barrier may be influenced when the WZ segments become shorter. In particular, the barrier could be lowered for shorter segments~\cite{chen_conduction_2017}. The observation made for the 80\,nm barrier suggests that the barrier emission effect levels out for large biases, which may be a sign of the onset of electron tunneling at larger biases. The shape of the resulting temperature distribution strongly suggests a spatially distributed Peltier heating/cooling power.
    
    \begin{figure}
        \centering
        \includegraphics[width=\linewidth]{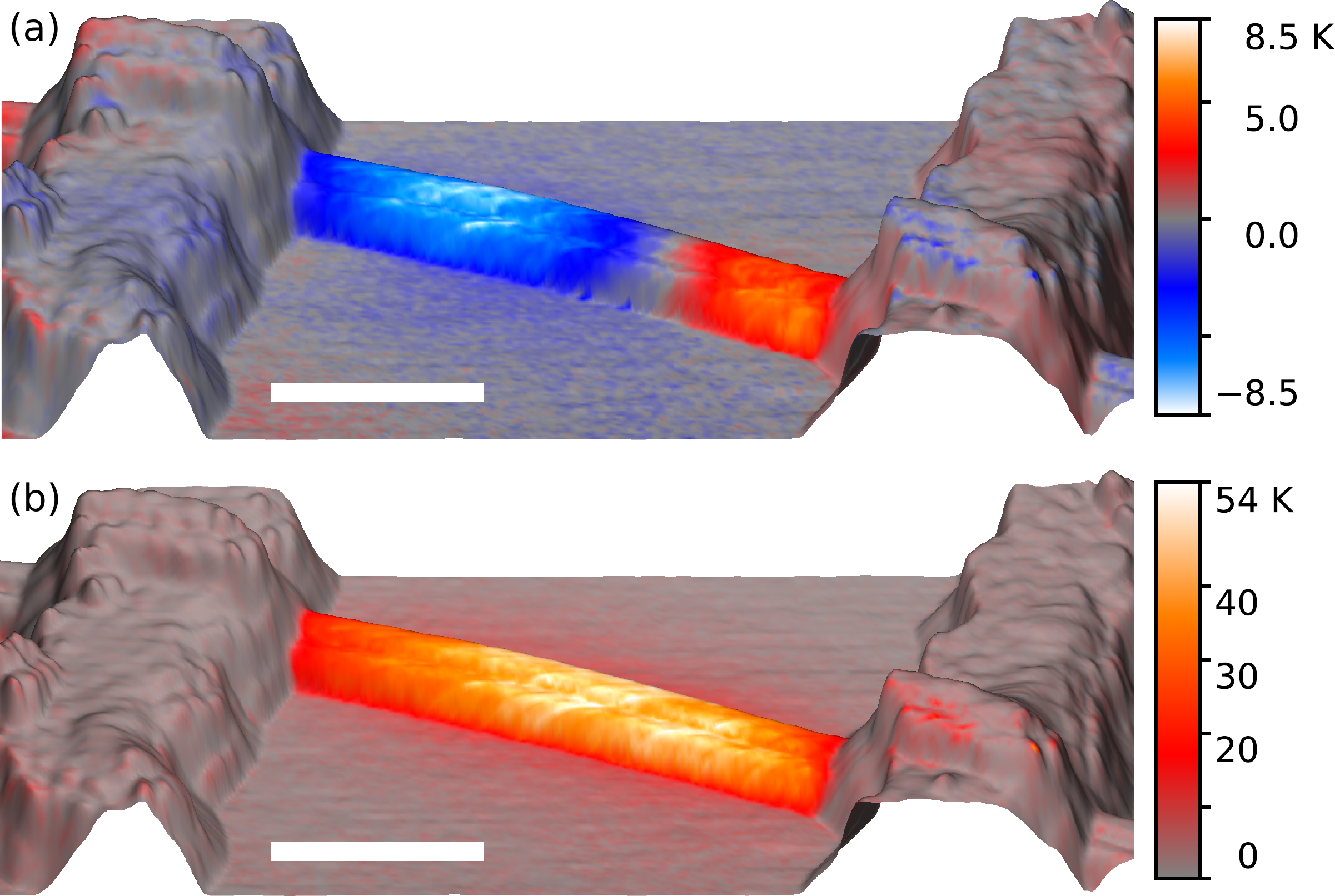}
        \caption{Data obtained for a fourth device, again with a target WZ segment length of 80\,nm. A slower, more detailed scan at $I = 45\,\mu$A is shown (see experimental section for details). Shown are 3D rendered overlay maps of AFM topography with (a) $\dT\subtxt{Peltier}$, (b) $\dT\subtxt{Joule}$. Orthographic projections. The fast scanning direction is horizontal. Scale bars 300\,nm.}
        \label{fig:wire5_3d_together}
    \end{figure}
    
\begin{figure}
        \centering
        \includegraphics[width=\linewidth]{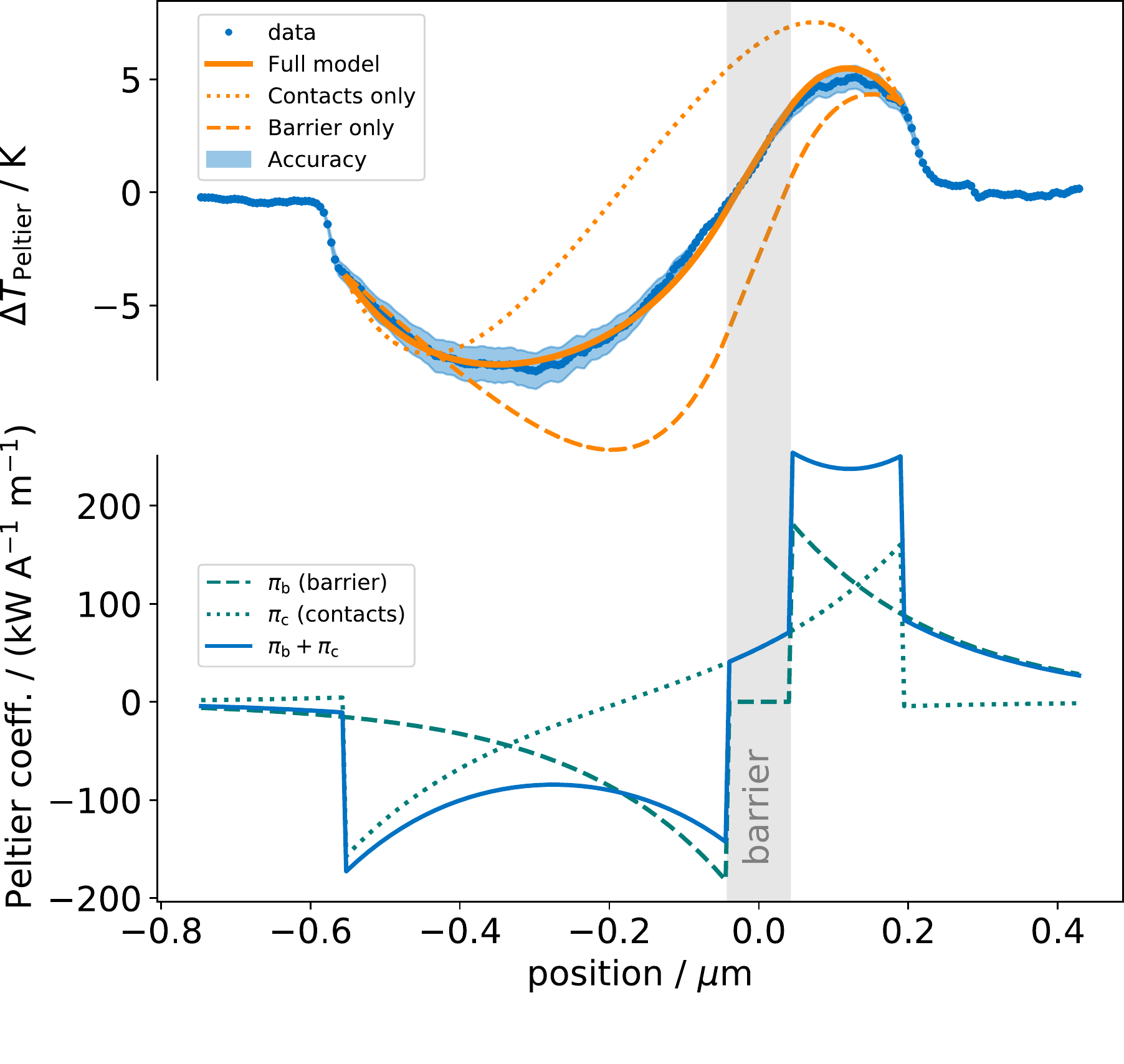}
        \caption{(Top) Line profile of $\dT\subtxt{Peltier}$ along the nanowire from \ref{fig:wire5_3d_together}. Averaging width of 20 pixels at a pixel size of 2.5\,nm. A fit of \ref{eqn:heat_equation_pelt} is drawn, together with a model that tries to explain all observed Peltier power with only the contact Peltier effect, and a solution which tries to explain all observed Peltier power is with only the barrier Peltier effect. The blue shaded area indicates the accuracy limits. See supporting information for details. (Bottom) Electron thermalization model. The Peltier coefficient per unit length is decaying exponentially away from junctions. The electron thermalization length is comparable to the system size.}
        \label{fig:wire5_fit_along_together}
    \end{figure}

    The qualitative discussion above calls for a more quantitative analysis. To this end, we have investigated a fourth device, again with a nominal WZ segment length of 80\,nm. We performed a scan at a modulation current amplitude of $I = 45\,\mu$A, but this time focusing on maximizing resolution and sensitivity with smaller pixel size and a slower modulation frequency (see experimental section for details). The result is shown in \ref{fig:wire5_3d_together}. Again, the transition between heating and cooling is not centered between the contacts. A line profile for $\dT\subtxt{Peltier}$ along the wire is shown in \ref{fig:wire5_fit_along_together}. The exact position of this profile on the temperature map is shown in the supporting information.
    
    Considering that phonons are the dominant carriers of heat in InAs NWs, and that the phonon mean free path is limited by the NW diameter~\cite{swinkels_diameter_2015}, a one-dimensional heat diffusion model is justified. Note that the charge carriers have significantly longer equilibration lengths, on the order of hundreds of nanometers, in comparison to phonons limited to the wire diameter.
    As noted above, the Peltier power is spatially distributed. We therefore introduce a Peltier coefficient per unit length (unit W/A/m) and denote it with $\pi$. We then consider the diffusion of heat from Peltier sources and sinks independently from Joule heating (see appendix for details).  All measurements and modulations are made on time scales much longer than the equilibration time scale of the nanowire system. We can thus regard the steady-state solution of the one-dimensional heat diffusion equation for $\dT\subtxt{Peltier}$:
    \begin{align}
                \kappa A \frac{\partial^{2} \dT\subtxt{Peltier}}{\partial x^{2}}(x) = - \pi(x) I + g \dT\subtxt{Peltier}(x)~,\label{eqn:heat_equation_pelt}
    \end{align}
    where the $x$-direction is along the wire axis, $\kappa$ is the thermal conductivity, $A$ is the cross sectional area of the wire, $\pi(x)$ is the previously mentioned Peltier coefficient per unit length, and $g$ is the thermal conductance per unit length from nanowire to underlying substrate.   
    This model predicts $\dT\subtxt{Peltier}(x)$, which we will compare with an experimental line profile taken from an SThM map of $\dT\subtxt{Peltier}$ along the wire. 
    
    The physical properties of the system extracted as fit parameters are given in \ref{tab:params}. The model has seven parameters that could be extracted by the fitting. Owing to the high spatial resolution of the experiment, the model has 143 remaining degrees of freedom. While the per-pixel temperature sensitivity of approximately 250\,mK ($\approx$56\,mK for an averaged point on a profile line), is determined by random noise, the accuracy of a given temperature value on the order of 10\% is determined by probe variations, underlying assumptions and the sensor calibration. The range within the limits of accuracy are drawn in light blue in \ref{fig:wire5_fit_along_together}. A discussion of the uncertainties can be found in the supporting information. See appendix for details of the fitting procedure.
    
    A central aspect of this analysis is the shape of $\pi(x)$. As discussed above, we expect a significant distribution of $\pi(x)$. As this distribution originates from the time or distance required for non-equilibrium charge carriers to equilibrate with the lattice through inelastic scattering processes, we expect $\pi(x)$ to have the shape of an exponential decay away from junctions, where contributions from different junctions are added. Simulations of electron-phonon energy exchange at thin energy barriers in a semiconductor material have shown that the power density in fact follows an exponential decay in essence~\cite{lake_energy_1992}, with equal and opposite magnitudes on either side of the barrier. Since we have the same material (ZB InAs) in which charge carriers are injected/extracted at the contacts, it is reasonable to assume the same decay length for the contact Peltier effect. We denote the electron thermalization length with $\Lambda$ and design a parametrization of $\pi(x)$ around these insights, which is illustrated in the bottom graph of \ref{fig:wire5_fit_along_together} for the optimal set of parameters found. We note that the description of the source term $\pi(x)$ relates to the electron system only and does not contain contributions from phonons, which enter the thermal conductivity terms. We further note, that this description is essentially different from a spatially distributed Seebeck coefficient which can be related to carrier denisties around a p-n junction in semiconductors\cite{Lyeo2004}.
    
    \renewcommand{\arraystretch}{1.2}

    \begin{table}
        \caption{Physical parameters extracted from fit}
        \label{tab:params}
        \centering
        \begin{tabular}{@{}lcl@{}}
            \hline
            Parameter                       &   {Value}     &   {Unit}                              \\
            \hline\hline
            $x\subtxt{b}$                   &   $0 \pm 3$      &   nm                                 \\
            $l\subtxt{b}$                   &   $84 \pm 4$      &  nm                                 \\
            $\Lambda$                       &   $223 \pm 9$     &  nm                                 \\
            $\kappa$                        &   $8.9 \pm 0.5$       &   W/m/K                                \\
            $\pi_{0,\textrm{b}}$            &   $39 \pm 7$    &   kW/(Am)     \\
            $\pi_{0,\textrm{c}}$            &   $35 \pm 5$     &   W/(Am)     \\
            $g$                             &   $0.3 \pm 0.1$      &   W/m/K  \\
            \hline
        \end{tabular}
    \end{table}
    
    Let us discuss the values of the physical parameters we have extracted:
    
    The barrier length (WZ segment length) $l\subtxt{b}$ and the barrier location $x\subtxt{b}$: In an earlier study, the actual WZ segment length achieved when targeting $80\,nm$ was determined to be statistical with a value of $82 \pm 17$\,nm, which is consistent with the value for $l\subtxt{b}$ that we find here.As expected from the tolerances of the fabrication process, the most plausible location of the barrier is not right in between the contacts, here set to zero. 
    
    Thermoelectric properties: 
    The effective Peltier coefficient $\Pi$ associated with a junction (unit W/A) can be found by integration of the according component of $\pi$. For our exponential decay model, this simply means $\Pi = \Lambda\pi$. We find a value associated with the barrier of $\Pi\subtxt{b} = 39 \pm 7$\,mV. The Peltier coefficient and the Seebeck coefficient $S$ are fundamentally connected through $S = \Pi / T\subtxt{avg}$. For the average temperature $T\subtxt{avg}$, we take room temperature plus the average value of $\dT\subtxt{Joule}$ on the wire. We obtain a value for the  barrier Seebeck coefficient of $S\subtxt{b} = 89 \pm 21\,\mu$W/K. As the value of $S$ depends on the chemical potential of the sample, which is in this case prone to unintentional doping and charging of the NWs, it is not straightforward to compare this to literature values. Nevertheless, it is interesting to make a comparison to previously measured intrinsic Seebeck coefficient of an InAs nanowire\cite{karg_full_2014}, which is about half in magnitude compared to the value we find here.
    \ref{fig:wire5_fit_along_together} also shows solutions to \ref{eqn:heat_equation_pelt} in which all heat pumping power is purely due to the contacts, and purely due to the barrier emission effect, respectively. If only the contacts contributed, the profile would be symmetric in between the contacts. If the effect was purely due to the barrier, the asymmetry would be even more pronounced, and the peaks of $\dT\subtxt{Peltier}$ would lie closer to the presumed barrier location. The observed behavior can only be explained when taking into account both effects.
    
    The value for thermal conductivity $\kappa$ of the InAs nanowire we find here is consistent with the  experimental results: Values for $\kappa$ of single crystalline InAs nanowires of both the ZB and the WZ phase have been determined experimentally to lie in a range of 8 to 10\,W/m/K at room temperature for similar nanowire diameters~\cite{zhou_thermal_2011}. Lower values were reported for wires in which a high defect density was either confirmed~\cite{karg_full_2014}, or could be expected~\cite{swinkels_diameter_2015}. For these wires, a value for $\kappa$ around 3\,W/m/K was measured. We expect our wires to be largely defect-free except for a few twin defects and the intentionally introduced WZ segment~\cite{chen_conduction_2017} and therefore also expected $\kappa$ to be close to the values determined for single-crystalline samples.
    
    The thermoelectric figure of merit $ZT = \frac{S^{2}T\subtxt{avg}}{\rho\kappa}$ is often used to evaluate the potential of a thermoelectric system for technical applications~\cite{snyder_complex_2010}. From the device's current-voltage characteristics, we estimate $\rho = 22 \pm 6\,\mu\Omega$m, yielding $ZT = 0.02 \pm 0.01$ for the barrier system.
    
    Thermalization length $\Lambda$: The decay length is related to the inelastic scattering length of the emitted electrons. The value we find is on the same order as values reported for the mean free path defined in the context of charge carrier mobility~\cite{chuang_ballistic_2013, zhou_scanned_2006, chen_thermoelectric_2018}. It is however important to note that although the two quantities are related through a notion of scattering length, only a limited comparison is possible.
    
    %Elastic scattering does for instance not contribute to thermalization, but it does have an impact on charge carrier mobility. On the other hand, the thermalization length can be affected by electron-electron scattering, which is not important for the mean free path associated with mobility, because it conserves the total momentum of the electron system.
    
    Substrate coupling $g$ (thermal conductance per unit length between nanowire and substrate): The thermal conductance between nanowire and underlying substrate, which is a parameter in our model, is not easy to extract from other experiments\cite{cahill_nanoscale_2002}. The value we extract is $0.3 \pm 0.1$\,W/m/K. For a given length, this conductance is governed  by two thermal resistances in series: The thermal boundary resistance $\R\subtxt{bnd}$ between the NW and the substrate material, and the thermal spreading resistance in the substrate, $\R\subtxt{spr}$. Subtracting the influence of the spreading resistance (details in the supporting information), we find a value for the interfacial thermal conductance of $G\subtxt{bnd} = 12.6 \pm 1.6$\,W/m/K.
    
\begin{figure}
        \centering
        \includegraphics[width=\linewidth]{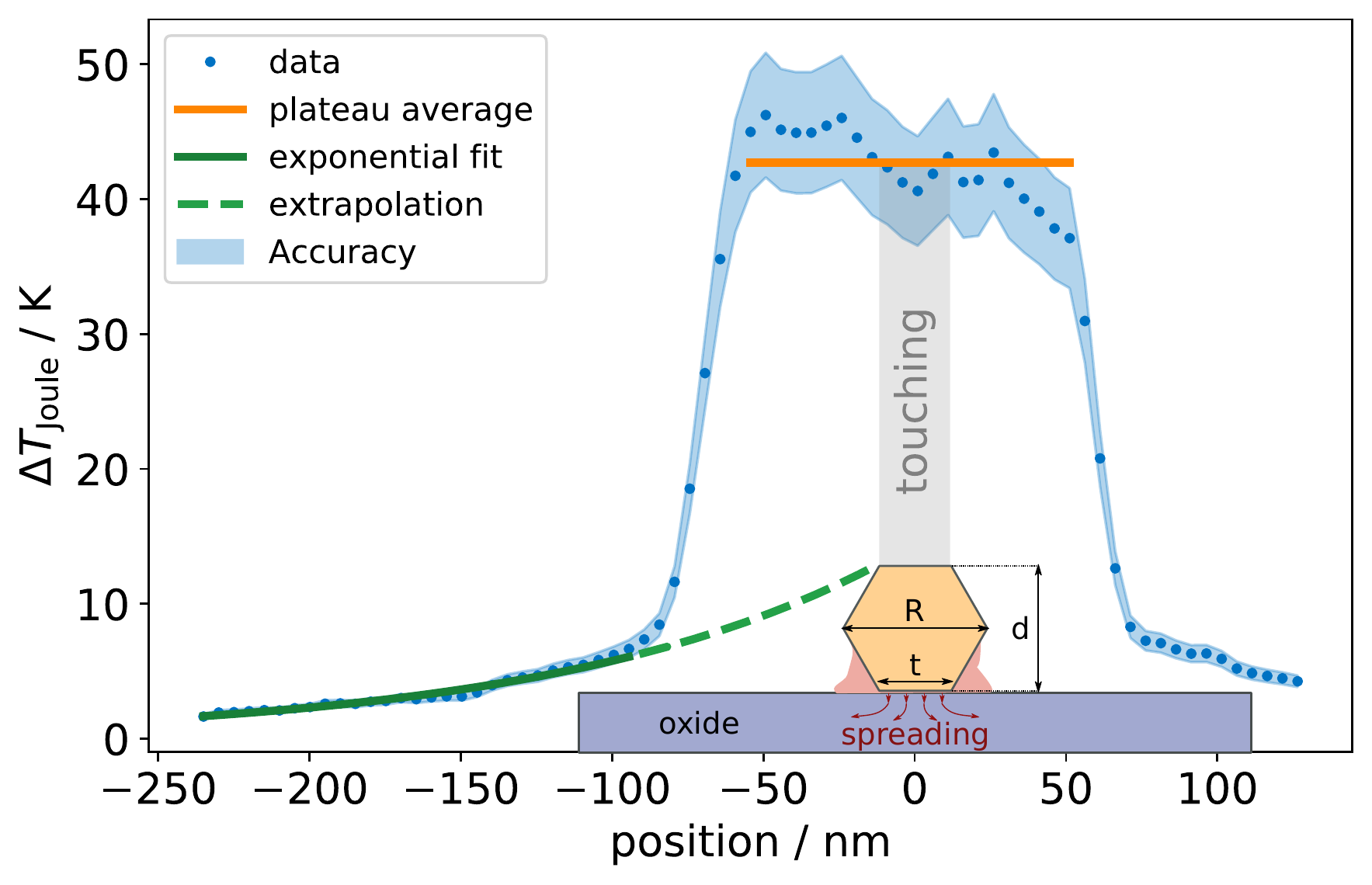}
        \caption{Joule heating temperature rise profile orthogonal to the wire. An exponential temperature decay away from the wire was fitted to points on the substrate. For geometrical reasons, the substrate temperature right next to the wire was not accessible in the measurement. The exponential trend was therefore extrapolated to obtain a value for the temperature jump from wire to substrate, from which the boundary resistance was inferred. The comparison with the Peltier profile data showed that the interface width is probably effectively wider than the edge width $t$, possibly due of the presence of an underfill originating from device processing (illustrated in light red). An illustration of the geometry and heat spreading into the substrate is provided. The blue shaded area indicates the limits of accuracy. See supporting information for details.}
        \label{fig:wire5_across_joule_fit}
    \end{figure}

    As a cross check, we determine $G\subtxt{bnd}$ by using the $\dT\subtxt{Joule}$ map and analyzing a profile line orthogonal to the nanowire axis, which is shown in \ref{fig:wire5_across_joule_fit}. The exact location of this profile line on the temperature map is shown in the supporting information. Heat spreading into the substrate is manifested as an exponential decay of $\dT\subtxt{Joule}$ away from the wire axis. The value of $\dT\subtxt{Joule}$ directly next to the wire is not accessible in our measurement because of the tip geometry. See \ref{fig:wire5_across_joule_fit} for an illustration of this issue. We fitted an exponential function to the accessible part of the heat spreading and extrapolated the temperature right next to the wire. The total thermal boundary conductance can now be estimated from the jump in $\dT\subtxt{Joule}$ to the value we measure on top of the wire, which is visible as a plateau in the temperature profile. (The plateau appears because the temperature within the nanowire does not vary over its cross-section, as expected from the large mean-free-path of phonons.) Assumptions about the interface geometry are needed to extract the thermal boundary conductance per unit area. The nanowire under investigation has a hexagonal cross-section. We denote the edge-to-edge distance with $d$, the corner-to-corner distance with $R$ and the edge length with $t$ (see illustration in \ref{fig:wire5_across_joule_fit}. Using $t$ as a lower limit for the interface width, we obtain a value of $G\subtxt{bnd} = 22 \pm 8$\,W/K/m$^2$.
    It is however likely that the actual contact area between wire and substrate is wider than $t$, for instance because there exists an underfill originating from the processing (such as electron beam resist residues), illustrated in light red in \ref{fig:wire5_across_joule_fit}. As an upper limit of the contact width, we extrapolate the touching point to the corner-to-corner width $R$ instead of the edge width $t$ and end up with a value of $14 \pm 5$\,W/K/m$^2$. This latter value is in good agreement with the value obtained from the Peltier fit. That the two different approaches yield similar results is seen as a confirmation that the method can in fact be used to extract values for the usually hard to characterize thermal boundary conductance associated with nanoscale thermal interfaces.
    
\section{Conclusion}
    In conclusion, we have observed thermoelectric effects around energy barriers formed by crystal phase engineering in InAs nanowires. Direct imaging of the resulting temperature fields with nanoscale spatial resolution was achieved. It was found that the thermoelectric heating/cooling power density is spatially distributed on a length scale comparable to the system size. The high spatial resolution and quantitative nature of the method allowed us to compare the measurement data with spatially dependent heat diffusion models. We were able to extract the thermalization length of non-equilibrium charge carriers. Thermal transport parameters, including thermal conductivity and thermal boundary resistance to the substrate, were quantified. For the latter there exists no other characterization method on this length scale. It was shown that the approach allows for a complete thermoelectic characterization. A thermoelectric figure of merit of $ZT = 0.02 \pm 0.01$ was calculated from the extracted parameters.
    
\section{Appendix}
    \subsection{Sample fabrication}
        The InAs nanowires were grown from Au aerosol seed particles on (111)B InAs substrates by low pressure metal-organic vapor phase epitaxy. The InAs crystal phase was controlled by changing the effective group V hydride flow, as described in refs.~\cite{ lehmann_general_2013, chen_conduction_2017, nilsson_single-electron_2016}.
        Resulting NW diameters were around $60 \pm 5$\,nm. Electron beam lithography was used to define the contact area, which was then etched in a mixture of (NH$_4$)$_2$S$_x$ and H$_2$O 1:20 for 1\,min at 40\,$^\circ$C. Afterwards, a film of 25\,nm Ni and 75\,nm Au was evaporated and then lifted off in acetone.

    \subsection{Scanning probe thermometry}
        We again refer to the schematic of the method shown in \ref{fig:figure_1}b and provide some more details on the experimental method:
        A differential electrical signal is created with the help of a Wheatstone bridge circuit. The signal is then amplified, and for each pixel, both time-averaged and demodulated at $1f\subtxt{mod}$ and $2f\subtxt{mod}$. The resulting time-averaged signal $V\subtxt{DC}$ and the de-modulated amplitudes $V\subtxt{AC}$ at $1f\subtxt{mod}$ and $2f\subtxt{mod}$ then relate the according harmonic content of $T\subtxt{sample}$ to the known out-of-contact sensor temperature $\del{T}\subtxt{sensor, ooc}$ through~\cite{menges_nanoscale_2016, koonemann_nanoscale_2018}
        \begin{align}
                \del{T}\subtxt{sample} = \del{T}\subtxt{sensor, ooc} ~\times~ \frac{\delac{V}}{\deldc{V} - \delac{V}}~,
                \label{eqn:thermometry}
        \end{align}
        where $\deldc{V}$ is the time-averaged voltage drop over the Wheatstone bridge compared to when the tip is out of contact, and $\delac{V}$ is the de-modulated voltage amplitude at the momentarily considered harmonic of $f\subtxt{mod}$. For illustration, the raw signals for the high resolution scan are shown in the supporting information.
        The sensor is operated in an active mode, meaning that it also serves as a heater that creates a  ``thermal bias''. The out-of-contact sensor temperature was 274\,$^\circ$C as a response to an electrical power dissipation of 1.3\,mV in the sensor. We do not expect the heated tip to influence the temperature in the probed region beyond the uncertainties of the method, because the tip-sample thermal resistance is large compared to the thermal resistance inside the sample~\cite{menges_quantitative_2012}. Experiments were performed in high vacuum (1e-6\,mbar) in a electrically and acoustically shielded laboratory with temperature stabilization~\cite{lortscher_next-generation_2013}. As discussed elsewhere~\cite{konemann_thermal_2019} we do not expect thermal radiation to play a  role in the experiment, even though the structures are of sub-wavelength size~\cite{shen_surface_2009}. The dwell time is chosen such that the lock-in can integrate about 20 cycles of the signal. The sensor has a thermal time-constant of 38\,$\mu$s\cite{durig_fundamentals_2005}. Faster modulation therefore allows for faster scans, but the signals can get attenuated. This attenuation can be corrected when the time-constant is known, but it still leads to a lower signal-to-noise ratio~\cite{majumdar_scanning_1999}. For the low resolution measurements, $f\subtxt{mod}$ was 7234\,Hz and for the high resolution measurements, $f\subtxt{mod}$ was 1342\,Hz. Additionally, the high resolution scan has a four times smaller pixel size compared to the lower resolution series (2.5\,nm vs. 10\,nm).
        
    \subsection{Modelling methods}
        The steady state heat diffusion equation including both Peltier and Joule source terms is
        \begin{align}
                    \kappa A \frac{\partial^{2} \dT}{\partial x^{2}}(x) = -\rho(x) I^{2} - \pi(x) I + g \dT(x)~,\label{eqn:heat_equation}
        \end{align}
        where $\rho$ is electrical resistivity.
        The source term proportional to $I^2$ is Joule heating and the source term proportional to $I$ captures Peltier/thermoelectric effects.
        
        For the small temperature variations probed here, we may assume that none of the parameters is temperature, current, or voltage dependent. In this case, \ref{eqn:heat_equation} is fully linear and can be decomposed into equations that describe heat conduction for the two source terms individually, in accordance with the thermometry method employed, which can discern between temperature changes induced by Joule and Peltier heat sources and sinks. For the Peltier part, we then obtain \ref{eqn:heat_equation_pelt}.

        The properties $g$, $\kappa$, $\rho$ and $A$ are not expected to vary along the nanowire and are assumed to be independent of $x$.
        It has been predicted and experimentally shown~\cite{zhou_thermal_2011}, that the thermal impedance mismatch between WZ InAs and ZB InAs is small. This is why in the model, $\kappa$ is assumed to be the same for WZ and ZB. There are however experimental indications that a thermal resistance is associated with an interface between WZ and ZB~\cite{karg_full_2014, swinkels_diameter_2015}. This interfacial resistance has been estimated and included in the model. See supporting information for details.
        
        For extracting physical parameters, we solve \ref{eqn:heat_equation_pelt} numerically with a boundary value solver for ordinary differential equations and perform a weighted non-linear least square fit of the model parameters~\cite{oliphant_python_2007, strutz_data_2016}. A discussion of the uncertainties is provided in the supporting information.

    \begin{acknowledgements}
        The authors thank U. Drechsler, S. Reidt, and A. Zulji for technical assistance, A. Schenk, K. Moselund, W. Riess, and H. Riel for continuous support, L Gignac for the TEM image shown in the supporting information, and S. H\"onl for helping us with 3D rendering. This project has received funding from the European Union's Horizon 2020 research and innovation programme under Grant Agreement No. 766853 (EFINED) and No. 767187 (QuIET).
    \end{acknowledgements}
     \section{Supporting information} \label{sec:suppinfo}
         Peltier and Joule temperature change maps for all measured excitation amplitudes of the low-resolution series, estimation of WZ/ZB boundary resistance, raw data and 2D result plots with drawn profile lines for the high resolution measurement, discussion of uncertainties for the temperature profile analysis.

    \bibliography{zotero}

\begin{thebibliography}{45}
\expandafter\ifx\csname natexlab\endcsname\relax\def\natexlab#1{#1}\fi
\expandafter\ifx\csname bibnamefont\endcsname\relax
  \def\bibnamefont#1{#1}\fi
\expandafter\ifx\csname bibfnamefont\endcsname\relax
  \def\bibfnamefont#1{#1}\fi
\expandafter\ifx\csname citenamefont\endcsname\relax
  \def\citenamefont#1{#1}\fi
\expandafter\ifx\csname url\endcsname\relax
  \def\url#1{\texttt{#1}}\fi
\expandafter\ifx\csname urlprefix\endcsname\relax\def\urlprefix{URL }\fi
\providecommand{\bibinfo}[2]{#2}
\providecommand{\eprint}[2][]{\url{#2}}

\bibitem[{\citenamefont{Vining}(2009)}]{vining_inconvenient_2009}
\bibinfo{author}{\bibfnamefont{C.~B.} \bibnamefont{Vining}},
  \bibinfo{journal}{Nature Materials} \textbf{\bibinfo{volume}{8}},
  \bibinfo{pages}{83} (\bibinfo{year}{2009}), ISSN \bibinfo{issn}{1476-4660},
  \urlprefix\url{https://www.nature.com/articles/nmat2361}.

\bibitem[{\citenamefont{Shakouri and
  Bowers}(1997)}]{shakouri_heterostructure_1997}
\bibinfo{author}{\bibfnamefont{A.}~\bibnamefont{Shakouri}} \bibnamefont{and}
  \bibinfo{author}{\bibfnamefont{J.~E.} \bibnamefont{Bowers}},
  \bibinfo{journal}{Applied Physics Letters} \textbf{\bibinfo{volume}{71}},
  \bibinfo{pages}{1234} (\bibinfo{year}{1997}), ISSN \bibinfo{issn}{0003-6951},
  \urlprefix\url{https://aip.scitation.org/doi/abs/10.1063/1.119861}.

\bibitem[{\citenamefont{Benenti et~al.}(2017)\citenamefont{Benenti, Casati,
  Saito, and Whitney}}]{benenti_fundamental_2017}
\bibinfo{author}{\bibfnamefont{G.}~\bibnamefont{Benenti}},
  \bibinfo{author}{\bibfnamefont{G.}~\bibnamefont{Casati}},
  \bibinfo{author}{\bibfnamefont{K.}~\bibnamefont{Saito}}, \bibnamefont{and}
  \bibinfo{author}{\bibfnamefont{R.~S.} \bibnamefont{Whitney}},
  \bibinfo{journal}{Physics Reports} \textbf{\bibinfo{volume}{694}},
  \bibinfo{pages}{1} (\bibinfo{year}{2017}), ISSN \bibinfo{issn}{03701573},
  \urlprefix\url{https://linkinghub.elsevier.com/retrieve/pii/S0370157317301540}.

\bibitem[{\citenamefont{Horio and Yanai}(1990)}]{horio_numerical_1990}
\bibinfo{author}{\bibfnamefont{K.}~\bibnamefont{Horio}} \bibnamefont{and}
  \bibinfo{author}{\bibfnamefont{H.}~\bibnamefont{Yanai}},
  \bibinfo{journal}{IEEE Transactions on Electron Devices}
  \textbf{\bibinfo{volume}{37}}, \bibinfo{pages}{1093} (\bibinfo{year}{1990}),
  ISSN \bibinfo{issn}{0018-9383}.

\bibitem[{\citenamefont{Yang et~al.}(1993)\citenamefont{Yang, East, and
  Haddad}}]{yang_numerical_1993}
\bibinfo{author}{\bibfnamefont{K.}~\bibnamefont{Yang}},
  \bibinfo{author}{\bibfnamefont{J.~R.} \bibnamefont{East}}, \bibnamefont{and}
  \bibinfo{author}{\bibfnamefont{G.~I.} \bibnamefont{Haddad}},
  \bibinfo{journal}{Solid-State Electronics} \textbf{\bibinfo{volume}{36}},
  \bibinfo{pages}{321} (\bibinfo{year}{1993}), ISSN \bibinfo{issn}{00381101},
  \urlprefix\url{http://linkinghub.elsevier.com/retrieve/pii/0038110193900833}.

\bibitem[{\citenamefont{{Shakouri, A.} et~al.}(1998)\citenamefont{{Shakouri,
  A.}, {Lee, E.Y.}, {Smith, D.L.}, {Narayanamurti, V.}, {Bowers, J. E.}, and
  {Bowers, J. E.}}}]{shakouri_a._thermoelectric_1998}
\bibinfo{author}{\bibnamefont{{Shakouri, A.}}},
  \bibinfo{author}{\bibnamefont{{Lee, E.Y.}}},
  \bibinfo{author}{\bibnamefont{{Smith, D.L.}}},
  \bibinfo{author}{\bibnamefont{{Narayanamurti, V.}}},
  \bibinfo{author}{\bibnamefont{{Bowers, J. E.}}}, \bibnamefont{and}
  \bibinfo{author}{\bibnamefont{{Bowers, J. E.}}}, \bibinfo{journal}{Microscale
  Thermophysical Engineering} \textbf{\bibinfo{volume}{2}}, \bibinfo{pages}{37}
  (\bibinfo{year}{1998}), ISSN \bibinfo{issn}{1089-3954},
  \urlprefix\url{https://iom3.tandfonline.com/doi/abs/10.1080/108939598200097}.

\bibitem[{\citenamefont{Caroff et~al.}(2009)\citenamefont{Caroff, Dick,
  Johansson, Messing, Deppert, and Samuelson}}]{caroff_controlled_2009}
\bibinfo{author}{\bibfnamefont{P.}~\bibnamefont{Caroff}},
  \bibinfo{author}{\bibfnamefont{K.~A.} \bibnamefont{Dick}},
  \bibinfo{author}{\bibfnamefont{J.}~\bibnamefont{Johansson}},
  \bibinfo{author}{\bibfnamefont{M.~E.} \bibnamefont{Messing}},
  \bibinfo{author}{\bibfnamefont{K.}~\bibnamefont{Deppert}}, \bibnamefont{and}
  \bibinfo{author}{\bibfnamefont{L.}~\bibnamefont{Samuelson}},
  \bibinfo{journal}{Nature Nanotechnology} \textbf{\bibinfo{volume}{4}},
  \bibinfo{pages}{50} (\bibinfo{year}{2009}), ISSN \bibinfo{issn}{1748-3395},
  \urlprefix\url{https://www.nature.com/articles/nnano.2008.359}.

\bibitem[{\citenamefont{Lehmann et~al.}(2013)\citenamefont{Lehmann, Wallentin,
  Jacobsson, Deppert, and Dick}}]{lehmann_general_2013}
\bibinfo{author}{\bibfnamefont{S.}~\bibnamefont{Lehmann}},
  \bibinfo{author}{\bibfnamefont{J.}~\bibnamefont{Wallentin}},
  \bibinfo{author}{\bibfnamefont{D.}~\bibnamefont{Jacobsson}},
  \bibinfo{author}{\bibfnamefont{K.}~\bibnamefont{Deppert}}, \bibnamefont{and}
  \bibinfo{author}{\bibfnamefont{K.~A.} \bibnamefont{Dick}},
  \bibinfo{journal}{Nano Letters} \textbf{\bibinfo{volume}{13}},
  \bibinfo{pages}{4099} (\bibinfo{year}{2013}), ISSN \bibinfo{issn}{1530-6984},
  \urlprefix\url{https://doi.org/10.1021/nl401554w}.

\bibitem[{\citenamefont{Dick et~al.}(2010)\citenamefont{Dick, Caroff,
  Bolinsson, Messing, Johansson, Deppert, Wallenberg, and
  Samuelson}}]{dick_control_2010}
\bibinfo{author}{\bibfnamefont{K.~A.} \bibnamefont{Dick}},
  \bibinfo{author}{\bibfnamefont{P.}~\bibnamefont{Caroff}},
  \bibinfo{author}{\bibfnamefont{J.}~\bibnamefont{Bolinsson}},
  \bibinfo{author}{\bibfnamefont{M.~E.} \bibnamefont{Messing}},
  \bibinfo{author}{\bibfnamefont{J.}~\bibnamefont{Johansson}},
  \bibinfo{author}{\bibfnamefont{K.}~\bibnamefont{Deppert}},
  \bibinfo{author}{\bibfnamefont{L.~R.} \bibnamefont{Wallenberg}},
  \bibnamefont{and}
  \bibinfo{author}{\bibfnamefont{L.}~\bibnamefont{Samuelson}},
  \bibinfo{journal}{Semiconductor Science and Technology}
  \textbf{\bibinfo{volume}{25}}, \bibinfo{pages}{024009}
  (\bibinfo{year}{2010}), ISSN \bibinfo{issn}{0268-1242},
  \urlprefix\url{http://stacks.iop.org/0268-1242/25/i=2/a=024009}.

\bibitem[{\citenamefont{Pan et~al.}(2014)\citenamefont{Pan, Fu, Yu, Wang, Zhu,
  Nie, Wang, Chen, Xiong, von Moln{\'a}r et~al.}}]{pan_controlled_2014}
\bibinfo{author}{\bibfnamefont{D.}~\bibnamefont{Pan}},
  \bibinfo{author}{\bibfnamefont{M.}~\bibnamefont{Fu}},
  \bibinfo{author}{\bibfnamefont{X.}~\bibnamefont{Yu}},
  \bibinfo{author}{\bibfnamefont{X.}~\bibnamefont{Wang}},
  \bibinfo{author}{\bibfnamefont{L.}~\bibnamefont{Zhu}},
  \bibinfo{author}{\bibfnamefont{S.}~\bibnamefont{Nie}},
  \bibinfo{author}{\bibfnamefont{S.}~\bibnamefont{Wang}},
  \bibinfo{author}{\bibfnamefont{Q.}~\bibnamefont{Chen}},
  \bibinfo{author}{\bibfnamefont{P.}~\bibnamefont{Xiong}},
  \bibinfo{author}{\bibfnamefont{S.}~\bibnamefont{von Moln{\'a}r}},
  \bibnamefont{et~al.}, \bibinfo{journal}{Nano Letters}
  \textbf{\bibinfo{volume}{14}}, \bibinfo{pages}{1214} (\bibinfo{year}{2014}),
  ISSN \bibinfo{issn}{1530-6984},
  \urlprefix\url{https://doi.org/10.1021/nl4040847}.

\bibitem[{\citenamefont{Chen et~al.}(2017)\citenamefont{Chen, Lehmann, Nilsson,
  Kivisaari, Linke, Dick, and Thelander}}]{chen_conduction_2017}
\bibinfo{author}{\bibfnamefont{I.-J.} \bibnamefont{Chen}},
  \bibinfo{author}{\bibfnamefont{S.}~\bibnamefont{Lehmann}},
  \bibinfo{author}{\bibfnamefont{M.}~\bibnamefont{Nilsson}},
  \bibinfo{author}{\bibfnamefont{P.}~\bibnamefont{Kivisaari}},
  \bibinfo{author}{\bibfnamefont{H.}~\bibnamefont{Linke}},
  \bibinfo{author}{\bibfnamefont{K.~A.} \bibnamefont{Dick}}, \bibnamefont{and}
  \bibinfo{author}{\bibfnamefont{C.}~\bibnamefont{Thelander}},
  \bibinfo{journal}{Nano Letters} \textbf{\bibinfo{volume}{17}},
  \bibinfo{pages}{902} (\bibinfo{year}{2017}), ISSN \bibinfo{issn}{1530-6984},
  \urlprefix\url{https://doi.org/10.1021/acs.nanolett.6b04211}.

\bibitem[{\citenamefont{Belabbes et~al.}(2012)\citenamefont{Belabbes, Panse,
  Furthm{\"u}ller, and Bechstedt}}]{belabbes_electronic_2012}
\bibinfo{author}{\bibfnamefont{A.}~\bibnamefont{Belabbes}},
  \bibinfo{author}{\bibfnamefont{C.}~\bibnamefont{Panse}},
  \bibinfo{author}{\bibfnamefont{J.}~\bibnamefont{Furthm{\"u}ller}},
  \bibnamefont{and}
  \bibinfo{author}{\bibfnamefont{F.}~\bibnamefont{Bechstedt}},
  \bibinfo{journal}{Physical Review B} \textbf{\bibinfo{volume}{86}},
  \bibinfo{pages}{075208} (\bibinfo{year}{2012}),
  \urlprefix\url{https://link.aps.org/doi/10.1103/PhysRevB.86.075208}.

\bibitem[{\citenamefont{Nilsson et~al.}(2016)\citenamefont{Nilsson, Namazi,
  Lehmann, Leijnse, Dick, and Thelander}}]{nilsson_single-electron_2016}
\bibinfo{author}{\bibfnamefont{M.}~\bibnamefont{Nilsson}},
  \bibinfo{author}{\bibfnamefont{L.}~\bibnamefont{Namazi}},
  \bibinfo{author}{\bibfnamefont{S.}~\bibnamefont{Lehmann}},
  \bibinfo{author}{\bibfnamefont{M.}~\bibnamefont{Leijnse}},
  \bibinfo{author}{\bibfnamefont{K.~A.} \bibnamefont{Dick}}, \bibnamefont{and}
  \bibinfo{author}{\bibfnamefont{C.}~\bibnamefont{Thelander}},
  \bibinfo{journal}{Physical Review B} \textbf{\bibinfo{volume}{93}},
  \bibinfo{pages}{195422} (\bibinfo{year}{2016}),
  \urlprefix\url{https://link.aps.org/doi/10.1103/PhysRevB.93.195422}.

\bibitem[{\citenamefont{Nilsson et~al.}(2017)\citenamefont{Nilsson, Chen,
  Lehmann, Maulerova, Dick, and Thelander}}]{nilsson_parallel-coupled_2017}
\bibinfo{author}{\bibfnamefont{M.}~\bibnamefont{Nilsson}},
  \bibinfo{author}{\bibfnamefont{I.-J.} \bibnamefont{Chen}},
  \bibinfo{author}{\bibfnamefont{S.}~\bibnamefont{Lehmann}},
  \bibinfo{author}{\bibfnamefont{V.}~\bibnamefont{Maulerova}},
  \bibinfo{author}{\bibfnamefont{K.~A.} \bibnamefont{Dick}}, \bibnamefont{and}
  \bibinfo{author}{\bibfnamefont{C.}~\bibnamefont{Thelander}},
  \bibinfo{journal}{Nano Letters} \textbf{\bibinfo{volume}{17}},
  \bibinfo{pages}{7847} (\bibinfo{year}{2017}), ISSN \bibinfo{issn}{1530-6984},
  \urlprefix\url{https://doi.org/10.1021/acs.nanolett.7b04090}.

\bibitem[{\citenamefont{Akopian et~al.}(2010)\citenamefont{Akopian, Patriarche,
  Liu, Harmand, and Zwiller}}]{akopian_crystal_2010}
\bibinfo{author}{\bibfnamefont{N.}~\bibnamefont{Akopian}},
  \bibinfo{author}{\bibfnamefont{G.}~\bibnamefont{Patriarche}},
  \bibinfo{author}{\bibfnamefont{L.}~\bibnamefont{Liu}},
  \bibinfo{author}{\bibfnamefont{J.-C.} \bibnamefont{Harmand}},
  \bibnamefont{and} \bibinfo{author}{\bibfnamefont{V.}~\bibnamefont{Zwiller}},
  \bibinfo{journal}{Nano Letters} \textbf{\bibinfo{volume}{10}},
  \bibinfo{pages}{1198} (\bibinfo{year}{2010}), ISSN \bibinfo{issn}{1530-6984},
  \urlprefix\url{https://doi.org/10.1021/nl903534n}.

\bibitem[{\citenamefont{Wang et~al.}(2014)\citenamefont{Wang, Bi, Li, Long,
  Liu, Lv, Lu, Sun, and Liu}}]{wang_thermoelectric_2014}
\bibinfo{author}{\bibfnamefont{M.}~\bibnamefont{Wang}},
  \bibinfo{author}{\bibfnamefont{C.}~\bibnamefont{Bi}},
  \bibinfo{author}{\bibfnamefont{L.}~\bibnamefont{Li}},
  \bibinfo{author}{\bibfnamefont{S.}~\bibnamefont{Long}},
  \bibinfo{author}{\bibfnamefont{Q.}~\bibnamefont{Liu}},
  \bibinfo{author}{\bibfnamefont{H.}~\bibnamefont{Lv}},
  \bibinfo{author}{\bibfnamefont{N.}~\bibnamefont{Lu}},
  \bibinfo{author}{\bibfnamefont{P.}~\bibnamefont{Sun}}, \bibnamefont{and}
  \bibinfo{author}{\bibfnamefont{M.}~\bibnamefont{Liu}},
  \bibinfo{journal}{Nature Communications} \textbf{\bibinfo{volume}{5}},
  \bibinfo{pages}{4598} (\bibinfo{year}{2014}), ISSN \bibinfo{issn}{2041-1723},
  \urlprefix\url{https://www.nature.com/articles/ncomms5598}.

\bibitem[{\citenamefont{Yalon et~al.}(2017)\citenamefont{Yalon, Deshmukh, Rojo,
  Lian, Neumann, Xiong, and Pop}}]{yalon_spatially_2017}
\bibinfo{author}{\bibfnamefont{E.}~\bibnamefont{Yalon}},
  \bibinfo{author}{\bibfnamefont{S.}~\bibnamefont{Deshmukh}},
  \bibinfo{author}{\bibfnamefont{M.~M.} \bibnamefont{Rojo}},
  \bibinfo{author}{\bibfnamefont{F.}~\bibnamefont{Lian}},
  \bibinfo{author}{\bibfnamefont{C.~M.} \bibnamefont{Neumann}},
  \bibinfo{author}{\bibfnamefont{F.}~\bibnamefont{Xiong}}, \bibnamefont{and}
  \bibinfo{author}{\bibfnamefont{E.}~\bibnamefont{Pop}},
  \bibinfo{journal}{Scientific Reports} \textbf{\bibinfo{volume}{7}},
  \bibinfo{pages}{15360} (\bibinfo{year}{2017}), ISSN
  \bibinfo{issn}{2045-2322},
  \urlprefix\url{https://www.nature.com/articles/s41598-017-14498-3}.

\bibitem[{\citenamefont{Suh et~al.}(2010)\citenamefont{Suh, Kim, Kim, Kang,
  Lee, Khang, Park, Yoon, Im, and Ihm}}]{suh_thermoelectric_2010}
\bibinfo{author}{\bibfnamefont{D.-S.~.} \bibnamefont{Suh}},
  \bibinfo{author}{\bibfnamefont{C.~.} \bibnamefont{Kim}},
  \bibinfo{author}{\bibfnamefont{K.~H. P.~.} \bibnamefont{Kim}},
  \bibinfo{author}{\bibfnamefont{Y.-S.~.} \bibnamefont{Kang}},
  \bibinfo{author}{\bibfnamefont{T.-Y.~.} \bibnamefont{Lee}},
  \bibinfo{author}{\bibfnamefont{Y.~.} \bibnamefont{Khang}},
  \bibinfo{author}{\bibfnamefont{T.~S.~.} \bibnamefont{Park}},
  \bibinfo{author}{\bibfnamefont{Y.-G.~.} \bibnamefont{Yoon}},
  \bibinfo{author}{\bibfnamefont{J.~.} \bibnamefont{Im}}, \bibnamefont{and}
  \bibinfo{author}{\bibfnamefont{J.~.} \bibnamefont{Ihm}},
  \bibinfo{journal}{Applied Physics Letters} \textbf{\bibinfo{volume}{96}},
  \bibinfo{pages}{123115} (\bibinfo{year}{2010}), ISSN
  \bibinfo{issn}{0003-6951},
  \urlprefix\url{https://aip.scitation.org/doi/abs/10.1063/1.3259649}.

\bibitem[{\citenamefont{Lee et~al.}(2012)\citenamefont{Lee, Asheghi, and
  Goodson}}]{lee_impact_2012}
\bibinfo{author}{\bibfnamefont{J.}~\bibnamefont{Lee}},
  \bibinfo{author}{\bibfnamefont{M.}~\bibnamefont{Asheghi}}, \bibnamefont{and}
  \bibinfo{author}{\bibfnamefont{K.~E.} \bibnamefont{Goodson}},
  \bibinfo{journal}{Nanotechnology} \textbf{\bibinfo{volume}{23}},
  \bibinfo{pages}{205201} (\bibinfo{year}{2012}), ISSN
  \bibinfo{issn}{0957-4484},
  \urlprefix\url{https://doi.org/10.1088%2F0957-4484%2F23%2F20%2F205201}.

\bibitem[{\citenamefont{Pop}(2010)}]{pop_energy_2010}
\bibinfo{author}{\bibfnamefont{E.}~\bibnamefont{Pop}}, \bibinfo{journal}{Nano
  Research} \textbf{\bibinfo{volume}{3}}, \bibinfo{pages}{147}
  (\bibinfo{year}{2010}), ISSN \bibinfo{issn}{1998-0124, 1998-0000},
  \urlprefix\url{https://link.springer.com/article/10.1007/s12274-010-1019-z}.

\bibitem[{\citenamefont{Cahill et~al.}(2002)\citenamefont{Cahill, Ford,
  Goodson, Mahan, Majumdar, Maris, Merlin, and
  Phillpot}}]{cahill_nanoscale_2002}
\bibinfo{author}{\bibfnamefont{D.~G.} \bibnamefont{Cahill}},
  \bibinfo{author}{\bibfnamefont{W.~K.} \bibnamefont{Ford}},
  \bibinfo{author}{\bibfnamefont{K.~E.} \bibnamefont{Goodson}},
  \bibinfo{author}{\bibfnamefont{G.~D.} \bibnamefont{Mahan}},
  \bibinfo{author}{\bibfnamefont{A.}~\bibnamefont{Majumdar}},
  \bibinfo{author}{\bibfnamefont{H.~J.} \bibnamefont{Maris}},
  \bibinfo{author}{\bibfnamefont{R.}~\bibnamefont{Merlin}}, \bibnamefont{and}
  \bibinfo{author}{\bibfnamefont{S.~R.} \bibnamefont{Phillpot}},
  \bibinfo{journal}{Journal of Applied Physics} \textbf{\bibinfo{volume}{93}},
  \bibinfo{pages}{793} (\bibinfo{year}{2002}), ISSN \bibinfo{issn}{0021-8979},
  \urlprefix\url{https://aip.scitation.org/doi/10.1063/1.1524305}.

\bibitem[{\citenamefont{Menges et~al.}(2016{\natexlab{a}})\citenamefont{Menges,
  Mensch, Schmid, Riel, Stemmer, and Gotsmann}}]{menges_temperature_2016}
\bibinfo{author}{\bibfnamefont{F.}~\bibnamefont{Menges}},
  \bibinfo{author}{\bibfnamefont{P.}~\bibnamefont{Mensch}},
  \bibinfo{author}{\bibfnamefont{H.}~\bibnamefont{Schmid}},
  \bibinfo{author}{\bibfnamefont{H.}~\bibnamefont{Riel}},
  \bibinfo{author}{\bibfnamefont{A.}~\bibnamefont{Stemmer}}, \bibnamefont{and}
  \bibinfo{author}{\bibfnamefont{B.}~\bibnamefont{Gotsmann}},
  \bibinfo{journal}{Nature Communications} \textbf{\bibinfo{volume}{7}},
  \bibinfo{pages}{10874} (\bibinfo{year}{2016}{\natexlab{a}}), ISSN
  \bibinfo{issn}{2041-1723},
  \urlprefix\url{https://www.nature.com/articles/ncomms10874}.

\bibitem[{\citenamefont{K{\"o}onemann et~al.}(2018)\citenamefont{K{\"o}onemann,
  Vollmann, Menges, Chen, Ghazali, Yamaguchi, Ishibashi, Thelander, and
  Gotsmann}}]{koonemann_nanoscale_2018}
\bibinfo{author}{\bibfnamefont{F.}~\bibnamefont{K{\"o}onemann}},
  \bibinfo{author}{\bibfnamefont{M.}~\bibnamefont{Vollmann}},
  \bibinfo{author}{\bibfnamefont{F.}~\bibnamefont{Menges}},
  \bibinfo{author}{\bibfnamefont{I.}~\bibnamefont{Chen}},
  \bibinfo{author}{\bibfnamefont{N.~M.} \bibnamefont{Ghazali}},
  \bibinfo{author}{\bibfnamefont{T.}~\bibnamefont{Yamaguchi}},
  \bibinfo{author}{\bibfnamefont{K.}~\bibnamefont{Ishibashi}},
  \bibinfo{author}{\bibfnamefont{C.}~\bibnamefont{Thelander}},
  \bibnamefont{and} \bibinfo{author}{\bibfnamefont{B.}~\bibnamefont{Gotsmann}},
  in \emph{\bibinfo{booktitle}{2018 24rd {International} {Workshop} on
  {Thermal} {Investigations} of {ICs} and {Systems} ({THERMINIC})}}
  (\bibinfo{year}{2018}), pp. \bibinfo{pages}{1--6}.

\bibitem[{\citenamefont{K{\"o}nemann et~al.}(2019)\citenamefont{K{\"o}nemann,
  Vollmann, Wagner, Mohd~Ghazali, Yamaguchi, Stemmer, Ishibashi, and
  Gotsmann}}]{konemann_thermal_2019}
\bibinfo{author}{\bibfnamefont{F.}~\bibnamefont{K{\"o}nemann}},
  \bibinfo{author}{\bibfnamefont{M.}~\bibnamefont{Vollmann}},
  \bibinfo{author}{\bibfnamefont{T.}~\bibnamefont{Wagner}},
  \bibinfo{author}{\bibfnamefont{N.}~\bibnamefont{Mohd~Ghazali}},
  \bibinfo{author}{\bibfnamefont{T.}~\bibnamefont{Yamaguchi}},
  \bibinfo{author}{\bibfnamefont{A.}~\bibnamefont{Stemmer}},
  \bibinfo{author}{\bibfnamefont{K.}~\bibnamefont{Ishibashi}},
  \bibnamefont{and} \bibinfo{author}{\bibfnamefont{B.}~\bibnamefont{Gotsmann}},
  \bibinfo{journal}{The Journal of Physical Chemistry C}
  \textbf{\bibinfo{volume}{123}}, \bibinfo{pages}{12460}
  (\bibinfo{year}{2019}), ISSN \bibinfo{issn}{1932-7447},
  \urlprefix\url{https://doi.org/10.1021/acs.jpcc.9b00692}.

\bibitem[{\citenamefont{Harzheim et~al.}(2018)\citenamefont{Harzheim, Spiece,
  Evangeli, McCann, Falko, Sheng, Warner, Briggs, Mol, Gehring
  et~al.}}]{harzheim_geometrically_2018}
\bibinfo{author}{\bibfnamefont{A.}~\bibnamefont{Harzheim}},
  \bibinfo{author}{\bibfnamefont{J.}~\bibnamefont{Spiece}},
  \bibinfo{author}{\bibfnamefont{C.}~\bibnamefont{Evangeli}},
  \bibinfo{author}{\bibfnamefont{E.}~\bibnamefont{McCann}},
  \bibinfo{author}{\bibfnamefont{V.}~\bibnamefont{Falko}},
  \bibinfo{author}{\bibfnamefont{Y.}~\bibnamefont{Sheng}},
  \bibinfo{author}{\bibfnamefont{J.~H.} \bibnamefont{Warner}},
  \bibinfo{author}{\bibfnamefont{G.~A.~D.} \bibnamefont{Briggs}},
  \bibinfo{author}{\bibfnamefont{J.~A.} \bibnamefont{Mol}},
  \bibinfo{author}{\bibfnamefont{P.}~\bibnamefont{Gehring}},
  \bibnamefont{et~al.}, \bibinfo{journal}{Nano Letters}
  \textbf{\bibinfo{volume}{18}}, \bibinfo{pages}{7719} (\bibinfo{year}{2018}),
  ISSN \bibinfo{issn}{1530-6984},
  \urlprefix\url{https://doi.org/10.1021/acs.nanolett.8b03406}.

\bibitem[{\citenamefont{Lake and Datta}(1992)}]{lake_energy_1992}
\bibinfo{author}{\bibfnamefont{R.}~\bibnamefont{Lake}} \bibnamefont{and}
  \bibinfo{author}{\bibfnamefont{S.}~\bibnamefont{Datta}},
  \bibinfo{journal}{Physical Review B} \textbf{\bibinfo{volume}{46}},
  \bibinfo{pages}{4757} (\bibinfo{year}{1992}),
  \urlprefix\url{https://link.aps.org/doi/10.1103/PhysRevB.46.4757}.

\bibitem[{\citenamefont{Chen et~al.}(2018)\citenamefont{Chen, Burke, Svilans,
  Linke, and Thelander}}]{chen_thermoelectric_2018}
\bibinfo{author}{\bibfnamefont{I.-J.} \bibnamefont{Chen}},
  \bibinfo{author}{\bibfnamefont{A.}~\bibnamefont{Burke}},
  \bibinfo{author}{\bibfnamefont{A.}~\bibnamefont{Svilans}},
  \bibinfo{author}{\bibfnamefont{H.}~\bibnamefont{Linke}}, \bibnamefont{and}
  \bibinfo{author}{\bibfnamefont{C.}~\bibnamefont{Thelander}},
  \bibinfo{journal}{Physical Review Letters} \textbf{\bibinfo{volume}{120}},
  \bibinfo{pages}{177703} (\bibinfo{year}{2018}),
  \urlprefix\url{https://link.aps.org/doi/10.1103/PhysRevLett.120.177703}.

\bibitem[{\citenamefont{Ryu~Cho et~al.}(2017)\citenamefont{Ryu~Cho, Rawlings,
  Wolf, Spieser, Bisig, Reidt, Sousa, Khanal, Jacobs, and
  Knoll}}]{ryu_cho_sub-10_2017}
\bibinfo{author}{\bibfnamefont{Y.~K.} \bibnamefont{Ryu~Cho}},
  \bibinfo{author}{\bibfnamefont{C.~D.} \bibnamefont{Rawlings}},
  \bibinfo{author}{\bibfnamefont{H.}~\bibnamefont{Wolf}},
  \bibinfo{author}{\bibfnamefont{M.}~\bibnamefont{Spieser}},
  \bibinfo{author}{\bibfnamefont{S.}~\bibnamefont{Bisig}},
  \bibinfo{author}{\bibfnamefont{S.}~\bibnamefont{Reidt}},
  \bibinfo{author}{\bibfnamefont{M.}~\bibnamefont{Sousa}},
  \bibinfo{author}{\bibfnamefont{S.~R.} \bibnamefont{Khanal}},
  \bibinfo{author}{\bibfnamefont{T.~D.~B.} \bibnamefont{Jacobs}},
  \bibnamefont{and} \bibinfo{author}{\bibfnamefont{A.~W.} \bibnamefont{Knoll}},
  \bibinfo{journal}{ACS Nano} \textbf{\bibinfo{volume}{11}},
  \bibinfo{pages}{11890} (\bibinfo{year}{2017}), ISSN \bibinfo{issn}{1936-0851,
  1936-086X}, \urlprefix\url{http://pubs.acs.org/doi/10.1021/acsnano.7b06307}.

\bibitem[{\citenamefont{Karg et~al.}(2014)\citenamefont{Karg, Troncale,
  Drechsler, Mensch, Kanungo, Schmid, Schmidt, Gignac, Riel, and
  Gotsmann}}]{karg_full_2014}
\bibinfo{author}{\bibfnamefont{S.~F.} \bibnamefont{Karg}},
  \bibinfo{author}{\bibfnamefont{V.}~\bibnamefont{Troncale}},
  \bibinfo{author}{\bibfnamefont{U.}~\bibnamefont{Drechsler}},
  \bibinfo{author}{\bibfnamefont{P.}~\bibnamefont{Mensch}},
  \bibinfo{author}{\bibfnamefont{P.~D.} \bibnamefont{Kanungo}},
  \bibinfo{author}{\bibfnamefont{H.}~\bibnamefont{Schmid}},
  \bibinfo{author}{\bibfnamefont{V.}~\bibnamefont{Schmidt}},
  \bibinfo{author}{\bibfnamefont{L.}~\bibnamefont{Gignac}},
  \bibinfo{author}{\bibfnamefont{H.}~\bibnamefont{Riel}}, \bibnamefont{and}
  \bibinfo{author}{\bibfnamefont{B.}~\bibnamefont{Gotsmann}},
  \bibinfo{journal}{Nanotechnology} \textbf{\bibinfo{volume}{25}},
  \bibinfo{pages}{305702} (\bibinfo{year}{2014}), ISSN
  \bibinfo{issn}{0957-4484},
  \urlprefix\url{https://doi.org/10.1088%2F0957-4484%2F25%2F30%2F305702}.

\bibitem[{\citenamefont{Menges et~al.}(2016{\natexlab{b}})\citenamefont{Menges,
  Motzfeld, Schmid, Mensch, Dittberner, Karg, Riel, and
  Gotsmann}}]{menges_local_2016}
\bibinfo{author}{\bibfnamefont{F.}~\bibnamefont{Menges}},
  \bibinfo{author}{\bibfnamefont{F.}~\bibnamefont{Motzfeld}},
  \bibinfo{author}{\bibfnamefont{H.}~\bibnamefont{Schmid}},
  \bibinfo{author}{\bibfnamefont{P.}~\bibnamefont{Mensch}},
  \bibinfo{author}{\bibfnamefont{M.}~\bibnamefont{Dittberner}},
  \bibinfo{author}{\bibfnamefont{S.}~\bibnamefont{Karg}},
  \bibinfo{author}{\bibfnamefont{H.}~\bibnamefont{Riel}}, \bibnamefont{and}
  \bibinfo{author}{\bibfnamefont{B.}~\bibnamefont{Gotsmann}}, in
  \emph{\bibinfo{booktitle}{2016 {IEEE} {International} {Electron} {Devices}
  {Meeting} ({IEDM})}} (\bibinfo{year}{2016}{\natexlab{b}}), pp.
  \bibinfo{pages}{15.8.1--15.8.4}.

\bibitem[{\citenamefont{Rosa}(2012)}]{rosa_fundamentals_2012}
\bibinfo{author}{\bibfnamefont{A.~V.~D.} \bibnamefont{Rosa}},
  \emph{\bibinfo{title}{Fundamentals of {Renewable} {Energy} {Processes}}}
  (\bibinfo{publisher}{Academic Press}, \bibinfo{year}{2012}), ISBN
  \bibinfo{isbn}{978-0-12-397219-4}, \bibinfo{note}{google-Books-ID:
  7LX3yrVEgCMC}.

\bibitem[{\citenamefont{Swinkels et~al.}(2015)\citenamefont{Swinkels, Delft,
  Oliveira, Cavalli, Zardo, Heijden, and Bakkers}}]{swinkels_diameter_2015}
\bibinfo{author}{\bibfnamefont{M.~Y.} \bibnamefont{Swinkels}},
  \bibinfo{author}{\bibfnamefont{M.~R.~v.} \bibnamefont{Delft}},
  \bibinfo{author}{\bibfnamefont{D.~S.} \bibnamefont{Oliveira}},
  \bibinfo{author}{\bibfnamefont{A.}~\bibnamefont{Cavalli}},
  \bibinfo{author}{\bibfnamefont{I.}~\bibnamefont{Zardo}},
  \bibinfo{author}{\bibfnamefont{R.~W. v.~d.} \bibnamefont{Heijden}},
  \bibnamefont{and} \bibinfo{author}{\bibfnamefont{E.~P. A.~M.}
  \bibnamefont{Bakkers}}, \bibinfo{journal}{Nanotechnology}
  \textbf{\bibinfo{volume}{26}}, \bibinfo{pages}{385401}
  (\bibinfo{year}{2015}), ISSN \bibinfo{issn}{0957-4484},
  \urlprefix\url{https://doi.org/10.1088%2F0957-4484%2F26%2F38%2F385401}.

\bibitem[{\citenamefont{Lyeo et~al.}(2004)\citenamefont{Lyeo, Khajetoorians,
  Shi, Pipe, Ram, Shakouri, and Shih}}]{Lyeo2004}
\bibinfo{author}{\bibfnamefont{H.-K.} \bibnamefont{Lyeo}},
  \bibinfo{author}{\bibfnamefont{A.~A.} \bibnamefont{Khajetoorians}},
  \bibinfo{author}{\bibfnamefont{L.}~\bibnamefont{Shi}},
  \bibinfo{author}{\bibfnamefont{K.~P.} \bibnamefont{Pipe}},
  \bibinfo{author}{\bibfnamefont{R.~J.} \bibnamefont{Ram}},
  \bibinfo{author}{\bibfnamefont{A.}~\bibnamefont{Shakouri}}, \bibnamefont{and}
  \bibinfo{author}{\bibfnamefont{C.~K.} \bibnamefont{Shih}},
  \bibinfo{journal}{Science} \textbf{\bibinfo{volume}{303}},
  \bibinfo{pages}{816} (\bibinfo{year}{2004}).

\bibitem[{\citenamefont{Zhou et~al.}(2011)\citenamefont{Zhou, Moore, Bolinsson,
  Persson, Fr{\"o}berg, Pettes, Kong, Rabenberg, Caroff, Stewart
  et~al.}}]{zhou_thermal_2011}
\bibinfo{author}{\bibfnamefont{F.}~\bibnamefont{Zhou}},
  \bibinfo{author}{\bibfnamefont{A.~L.} \bibnamefont{Moore}},
  \bibinfo{author}{\bibfnamefont{J.}~\bibnamefont{Bolinsson}},
  \bibinfo{author}{\bibfnamefont{A.}~\bibnamefont{Persson}},
  \bibinfo{author}{\bibfnamefont{L.}~\bibnamefont{Fr{\"o}berg}},
  \bibinfo{author}{\bibfnamefont{M.~T.} \bibnamefont{Pettes}},
  \bibinfo{author}{\bibfnamefont{H.}~\bibnamefont{Kong}},
  \bibinfo{author}{\bibfnamefont{L.}~\bibnamefont{Rabenberg}},
  \bibinfo{author}{\bibfnamefont{P.}~\bibnamefont{Caroff}},
  \bibinfo{author}{\bibfnamefont{D.~A.} \bibnamefont{Stewart}},
  \bibnamefont{et~al.}, \bibinfo{journal}{Physical Review B}
  \textbf{\bibinfo{volume}{83}}, \bibinfo{pages}{205416}
  (\bibinfo{year}{2011}),
  \urlprefix\url{https://link.aps.org/doi/10.1103/PhysRevB.83.205416}.

\bibitem[{\citenamefont{Snyder and Toberer}(2010)}]{snyder_complex_2010}
\bibinfo{author}{\bibfnamefont{G.~J.} \bibnamefont{Snyder}} \bibnamefont{and}
  \bibinfo{author}{\bibfnamefont{E.~S.} \bibnamefont{Toberer}}, in
  \emph{\bibinfo{booktitle}{Materials for {Sustainable} {Energy}}}
  (\bibinfo{publisher}{Co-Published with Macmillan Publishers Ltd, UK},
  \bibinfo{year}{2010}), pp. \bibinfo{pages}{101--110}, ISBN
  \bibinfo{isbn}{978-981-4317-64-1},
  \urlprefix\url{https://www.worldscientific.com/doi/abs/10.1142/9789814317665_0016}.

\bibitem[{\citenamefont{Chuang et~al.}(2013)\citenamefont{Chuang, Gao, Kapadia,
  Ford, Guo, and Javey}}]{chuang_ballistic_2013}
\bibinfo{author}{\bibfnamefont{S.}~\bibnamefont{Chuang}},
  \bibinfo{author}{\bibfnamefont{Q.}~\bibnamefont{Gao}},
  \bibinfo{author}{\bibfnamefont{R.}~\bibnamefont{Kapadia}},
  \bibinfo{author}{\bibfnamefont{A.~C.} \bibnamefont{Ford}},
  \bibinfo{author}{\bibfnamefont{J.}~\bibnamefont{Guo}}, \bibnamefont{and}
  \bibinfo{author}{\bibfnamefont{A.}~\bibnamefont{Javey}},
  \bibinfo{journal}{Nano Letters} \textbf{\bibinfo{volume}{13}},
  \bibinfo{pages}{555} (\bibinfo{year}{2013}), ISSN \bibinfo{issn}{1530-6984},
  \urlprefix\url{https://doi.org/10.1021/nl3040674}.

\bibitem[{\citenamefont{Zhou et~al.}(2006)\citenamefont{Zhou, Dayeh, Aplin,
  Wang, and Yu}}]{zhou_scanned_2006}
\bibinfo{author}{\bibfnamefont{X.}~\bibnamefont{Zhou}},
  \bibinfo{author}{\bibfnamefont{S.~A.} \bibnamefont{Dayeh}},
  \bibinfo{author}{\bibfnamefont{D.}~\bibnamefont{Aplin}},
  \bibinfo{author}{\bibfnamefont{D.}~\bibnamefont{Wang}}, \bibnamefont{and}
  \bibinfo{author}{\bibfnamefont{E.~T.} \bibnamefont{Yu}},
  \bibinfo{journal}{Journal of Vacuum Science \& Technology B: Microelectronics
  and Nanometer Structures Processing, Measurement, and Phenomena}
  \textbf{\bibinfo{volume}{24}}, \bibinfo{pages}{2036} (\bibinfo{year}{2006}),
  ISSN \bibinfo{issn}{1071-1023},
  \urlprefix\url{https://avs.scitation.org/doi/full/10.1116/1.2213267}.

\bibitem[{\citenamefont{Menges et~al.}(2016{\natexlab{c}})\citenamefont{Menges,
  Riel, Stemmer, and Gotsmann}}]{menges_nanoscale_2016}
\bibinfo{author}{\bibfnamefont{F.}~\bibnamefont{Menges}},
  \bibinfo{author}{\bibfnamefont{H.}~\bibnamefont{Riel}},
  \bibinfo{author}{\bibfnamefont{A.}~\bibnamefont{Stemmer}}, \bibnamefont{and}
  \bibinfo{author}{\bibfnamefont{B.}~\bibnamefont{Gotsmann}},
  \bibinfo{journal}{Review of Scientific Instruments}
  \textbf{\bibinfo{volume}{87}}, \bibinfo{pages}{074902}
  (\bibinfo{year}{2016}{\natexlab{c}}), ISSN \bibinfo{issn}{0034-6748},
  \urlprefix\url{https://aip.scitation.org/doi/abs/10.1063/1.4955449}.

\bibitem[{\citenamefont{Menges et~al.}(2012)\citenamefont{Menges, Riel,
  Stemmer, and Gotsmann}}]{menges_quantitative_2012}
\bibinfo{author}{\bibfnamefont{F.}~\bibnamefont{Menges}},
  \bibinfo{author}{\bibfnamefont{H.}~\bibnamefont{Riel}},
  \bibinfo{author}{\bibfnamefont{A.}~\bibnamefont{Stemmer}}, \bibnamefont{and}
  \bibinfo{author}{\bibfnamefont{B.}~\bibnamefont{Gotsmann}},
  \bibinfo{journal}{Nano Letters} \textbf{\bibinfo{volume}{12}},
  \bibinfo{pages}{596} (\bibinfo{year}{2012}), ISSN \bibinfo{issn}{1530-6984},
  \urlprefix\url{https://doi.org/10.1021/nl203169t}.

\bibitem[{\citenamefont{L{\"o}rtscher et~al.}(2013)\citenamefont{L{\"o}rtscher,
  Widmer, and Gotsmann}}]{lortscher_next-generation_2013}
\bibinfo{author}{\bibfnamefont{E.}~\bibnamefont{L{\"o}rtscher}},
  \bibinfo{author}{\bibfnamefont{D.}~\bibnamefont{Widmer}}, \bibnamefont{and}
  \bibinfo{author}{\bibfnamefont{B.}~\bibnamefont{Gotsmann}},
  \bibinfo{journal}{Nanoscale} \textbf{\bibinfo{volume}{5}},
  \bibinfo{pages}{10542} (\bibinfo{year}{2013}),
  \urlprefix\url{https://pubs.rsc.org/en/content/articlelanding/2013/nr/c3nr03373b}.

\bibitem[{\citenamefont{Shen et~al.}(2009)\citenamefont{Shen, Narayanaswamy,
  and Chen}}]{shen_surface_2009}
\bibinfo{author}{\bibfnamefont{S.}~\bibnamefont{Shen}},
  \bibinfo{author}{\bibfnamefont{A.}~\bibnamefont{Narayanaswamy}},
  \bibnamefont{and} \bibinfo{author}{\bibfnamefont{G.}~\bibnamefont{Chen}},
  \bibinfo{journal}{Nano Letters} \textbf{\bibinfo{volume}{9}},
  \bibinfo{pages}{2909} (\bibinfo{year}{2009}), ISSN \bibinfo{issn}{1530-6984},
  \urlprefix\url{https://doi.org/10.1021/nl901208v}.

\bibitem[{\citenamefont{D{\"u}rig}(2005)}]{durig_fundamentals_2005}
\bibinfo{author}{\bibfnamefont{U.}~\bibnamefont{D{\"u}rig}},
  \bibinfo{journal}{Journal of Applied Physics} \textbf{\bibinfo{volume}{98}},
  \bibinfo{pages}{044906} (\bibinfo{year}{2005}), ISSN
  \bibinfo{issn}{0021-8979},
  \urlprefix\url{https://aip.scitation.org/doi/abs/10.1063/1.2006968}.

\bibitem[{\citenamefont{Majumdar}(1999)}]{majumdar_scanning_1999}
\bibinfo{author}{\bibfnamefont{A.}~\bibnamefont{Majumdar}},
  \bibinfo{journal}{Annual Review of Materials Science}
  \textbf{\bibinfo{volume}{29}}, \bibinfo{pages}{505} (\bibinfo{year}{1999}),
  ISSN \bibinfo{issn}{0084-6600},
  \urlprefix\url{https://www.annualreviews.org/doi/10.1146/annurev.matsci.29.1.505}.

\bibitem[{\citenamefont{Oliphant}(2007)}]{oliphant_python_2007}
\bibinfo{author}{\bibfnamefont{T.~E.} \bibnamefont{Oliphant}},
  \bibinfo{journal}{Computing in Science Engineering}
  \textbf{\bibinfo{volume}{9}}, \bibinfo{pages}{10} (\bibinfo{year}{2007}),
  ISSN \bibinfo{issn}{1521-9615}.

\bibitem[{\citenamefont{Strutz}(2016)}]{strutz_data_2016}
\bibinfo{author}{\bibfnamefont{T.}~\bibnamefont{Strutz}},
  \emph{\bibinfo{title}{Data {Fitting} and {Uncertainty}: {A} practical
  introduction to weighted least squares and beyond}}
  (\bibinfo{publisher}{Springer Vieweg}, \bibinfo{year}{2016}),
  \bibinfo{edition}{2nd} ed., ISBN \bibinfo{isbn}{978-3-658-11455-8},
  \urlprefix\url{https://www.springer.com/gp/book/9783658114558}.

\end{thebibliography}
    \bibliographystyle{apsrev}

\end{document}